\documentclass{iau-JDSS}
\usepackage{graphicx}

\title[SFR Calibration] 
{Calibration of Star-Formation Rate Measurements Across the Electromagnetic Spectrum}

\author[C.C. Popescu]   
{Cristina C. Popescu$^{1,2}$}
\affiliation{$^1$Jeremiah Horrocks Institute, University of Central Lancashire,
  PR1 2HE, Preston, UK \break email: cpopescu@uclan.ac.uk\\
$^2$ Scientific Visitor at the Max Planck Institut f\"ur Kernphysik,
Saupfercheckweg 1, 69117 Heidelberg, Germany\\[\affilskip]
}

\pubyear{2013}
\volume{Volume}  
\pagerange{1--7}
\date{?? and in revised form ??}
\jname{Highlights of Astronomy}
\editors{Thierry Montmerle, ed.}
\begin{document}

\maketitle

\section{SFR determinations from SED modelling}

During this meeting we have discussed a lot about different methods of deriving star
formation rates in galaxies and what the advantages and disadvantages of
using different calibrations are (see the comprehensive review talk by 
Veronique Buat at this meeting; see also review by Calzetti 2012). During the 
session on SFR determinations from 
SED modelling, but also throughout the meeting, two SED modelling
approaches have been presented and discussed: the energy balance method and the
radiative transfer modelling. 

The energy balance method relies on the
conservation of energy between the stellar light aborbed by dust and that
emitted (by the same dust) in the mid-IR/far-IR/submm. This is by far a superior
method to only fitting SED templates in a limited spectral range (see talk by
Denis Burgarella). Nonetheless, the energy-balance method is not a
self-consistent analysis, since the SED of dust emission is not calculated
according to the radiation fields originating from the stellar populations in
the galaxy under study, but rather according to some templates. The templates
can be either empirical (Xu et al. 1998, Devriendt et al. 1999,
Sajina et al. 2006, Marshall et al. 2007) or 
theoretical (Dale \& Helou 2002, Draine \& Li 2007, Natale et al. 2010). At
this meeting we saw applications of two energy balance methods, MAGPHYS (da
Cuhna et al. 2008) and CIGALE (Burgarella et al. 2005, Noll et al. 2009). The
energy balance method is a useful tool when dealing with large statistical
samples of galaxies for which little information is available regarding
morphology/type, orientation and overall size. This advantage comes
nevertheless with the disadvantage that the energy balance methods cannot take
into account the effect on the dust attenuation and therefore also on the dust
emission of the different geometries of stars and dust present in galaxies of
different morphological types, neither can it take into account the 
anisotropies in the predicted stellar light due to disk inclination 
(when a disk geometry is present). These methods can therefore only be used in
a statistical sense, when dealing with overall trends in galaxy populations. 

The radiative transfer method is the only one
that can self-consistently calculate the dust emission SEDs based on an
explicit calculation of the radiation fields heating the dust, consequently
derived from the attenuated stellar populations in the galaxy under study. At
this meeting we saw applications of the RT model of Popescu et al. (2011). 
This method can take advantage of the constraints provided by available optical
information like morphology, disk-to-bulge ratio, disk inclination (when a disk
morphology is present) and size. 
As opposed to the energy balance method, this advantage comes with the 
disatvantage that such information is not always easily available. Nonetheless
these radiation transfer methofs could perhaps be adapted to incorporate this
information (when missing) in the form of free parameters of the model, 
though such attempts have not yet been made. Another drawback of these methods 
was that radiative transfer calculations are notorious for being
computationally very time consuming, and as such, detailed calculations have
been mainly used for a small number of galaxies (Popescu et al. 2000,
Misiriotis et al. 2001, Popescu et al. 2004, Bianchi 2008, Baes et al. 2010, 
MacLachlan et al. 2011, Schechtman-Rook et al. 2012, de Looze et
al. 2012a,b). This situation has been recently changed, with the creation of
large libraries of radiative transfer model SEDs, as performed by Siebenmorgen
\& Krugel (2007) for starburst galaxies, Groves et al. (2008) for star-forming
regions/starburt galaxies and Popescu et al. (2011) for spiral
galaxies. 

Reviews on determination of star-formation in galaxies have so far not
included discussions on the use of radiative transfer methods, with the 
exception of the review of Kylafis \& Misiriotis (2006). With the new 
developments
resulting in the creation of libraries of RT models, we can now start to
include radiative models as main topics of discussion. Indeed, at this meeting  
 we emphasised that they are in fact the most realible 
way of deriving star formation rates in galaxies. In this way
the SFRs are derived self-consistently using information from the whole range 
of the electromagnetic spectrum, from the UV to the FIR/submm, incorporating
information with morphological constraints (primarily from optical imaging).
Here we did not 
consider radio and Xray emission, though these emissions have been discussed
in other sessions of this meeting (e.g. talk by Bret Lehmer). In particular the SED modelling has been
discussed in conjunction with the most difficult cases, namely those of
translucent galaxies: galaxies with both optically thin and thick
components. In one way optically thin galaxies are more easily dealt with,
since most of the information on SFR can be derived from the UV. Very 
optically thick cases are also easy from this point of view, since SFR can 
be derived from their FIR emission, providing one can isolate the AGN
powered emission. But the most difficult cases are the translucent
galaxies. These are essentially the spiral galaxies in the Local Universe, and
probably a large fraction of the star forming dwarf galaxies - which dominate
the population of galaxies in the Local Universe, and which also host most of
the star formation activity taking place in the local Universe. We showed in
this meeting how important it is to quantify this star 
formation activity. We also discussed how important it is to quantify the star
formation activity in the high redshift Universe; it is just that for the 
moment we do not have enough detailed information to be able to do the 
same type of analysis that we can now do for the Local Universe.
Here I will summarise the main points we addressed:\\

\begin{figure}
 \includegraphics[scale=0.5]{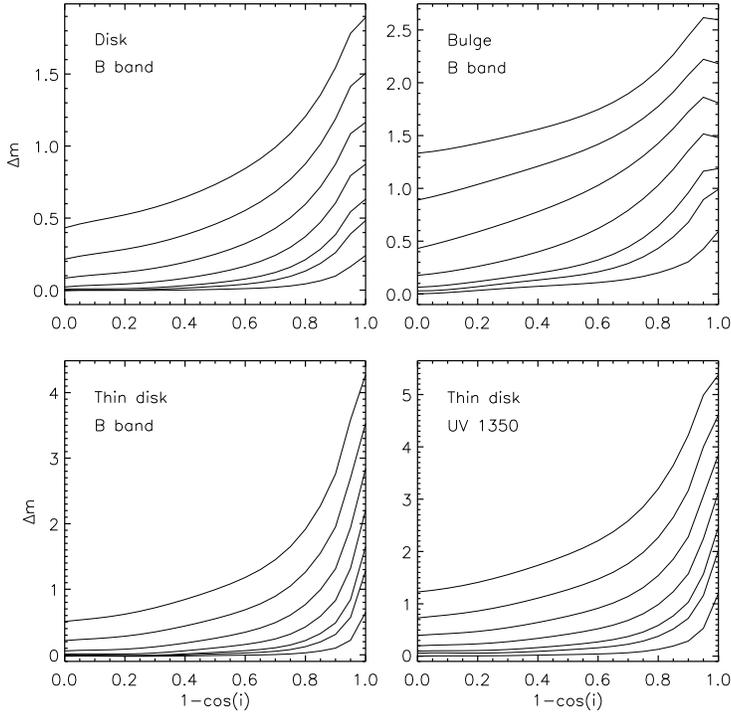}
  \caption{Predictions for the attenuation-inclination relation for
    different stellar components from Tuffs et
    al. (2004). From bottom to top the curves correspond to central face-on B
    band optical depth $\tau_B^f$ of 0.1,0.3,0.5,1.0,2.0,4.0,8.0}\label{fig:attenuation_incl_components}
\end{figure}

\begin{figure}
 \includegraphics[scale=0.6]{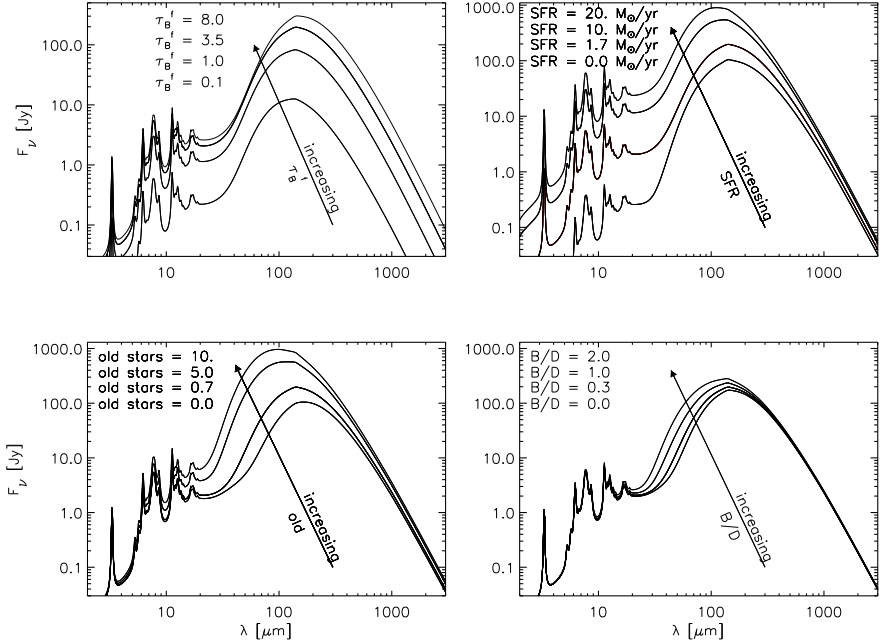}
  \caption{Predictions for dust and PAH emission SEDs based on the model of
    Popescu et al. (2011). In going clock-wise from the top-left, the different
    panels show the effect of changing the dust opacity, the luminosity of
    the young stellar populations (SFR), the luminosity of the old stellar
    populations (old) and the bulge-to-disk (B/D) ratio. In each panel only one
  parameter at a time is changed, while keeping the remaining ones fixed.}
\label{fig:sed}
\end{figure}

1. {\it Why is it so difficult to calibrate SFR in spiral galaxies and why do 
we need to follow the fate of photons with radiative transfer calculations?}
\begin{figure}
 \includegraphics[scale=0.4]{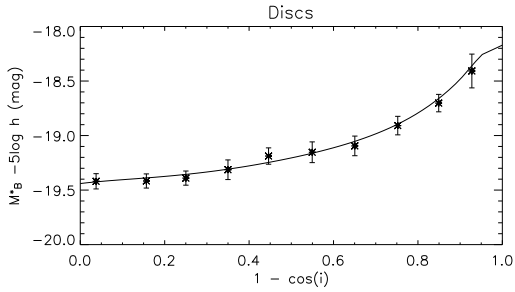}
  \caption{The attenuation-inclination relation from Driver et
    al. (2007). The symbols deliniate the empirical relation derived for disks
    from the Millenium Galaxy Survey while the solid line is the prediction
    from the model of Tuffs et al. (2004).}\label{fig:attenuation_incl_disk}
\end{figure}
\begin{itemize}
\item Fundamentally, any fixed observed luminosity in dust emission can be
powered either by a small fraction of a large quantity of optical light
from older stellar populations, or by a large fraction of a small quantity
of UV light from younger stellar populations.
Only radiation transfer techniques can unravel this dichotemy. 
\item Because of the disky nature of these systems, the direct UV and optical
  light is highly anisotropic, and the attenuation of the stellar photons will
  depend on the {\bf viewing angle} and {\bf wavelength}. 
  Fig.~\ref{fig:attenuation_incl_disk} illustrates the strong dependence
  of the observed luminosity of stellar disks on the viewing angle. 
\item Dust in galaxies has a very complex structure, containing both 
{\bf diffuse}  components on kiloparsec scales, as well as {\bf localised} 
components, at the pc scales, associated with the star forming regions. 
The escape of radiation from 
these two components is very different, as it is the heating of dust in the 
diffuse medium and in the  star-forming clouds.
\item Disk galaxies have different morphological components, in particular 
{\bf disks} and {\bf bulges}. The attenuation characteristics of disks is very 
different 
from those of bulges (see Fig.~\ref{fig:attenuation_incl_components}) and these need to be properly taken into account when 
dealing with the integrated emission from galaxies.
\item Different stellar populations have {\bf different spatial distributions} 
with 
respect to the dust distribution, and again their attenuation characteristics 
will differ, as will their contribution to heating the
dust. Fig.~\ref{fig:attenuation_incl_components} illustrates the different
behaviour of the variation of attenuation of light coming from 
different stellar populations with inclination and dust
opacity. Fig.~\ref{fig:sed} also shows how the dust and PAH emission SEDs are 
changed for various contributions coming from the old and young stellar
populations in the disk, as well as from the old stellar populations in the
bulge. \\
\end{itemize}

2. {\it Why do SFR calibrators work?}

\begin{figure}

\includegraphics[scale=0.38]{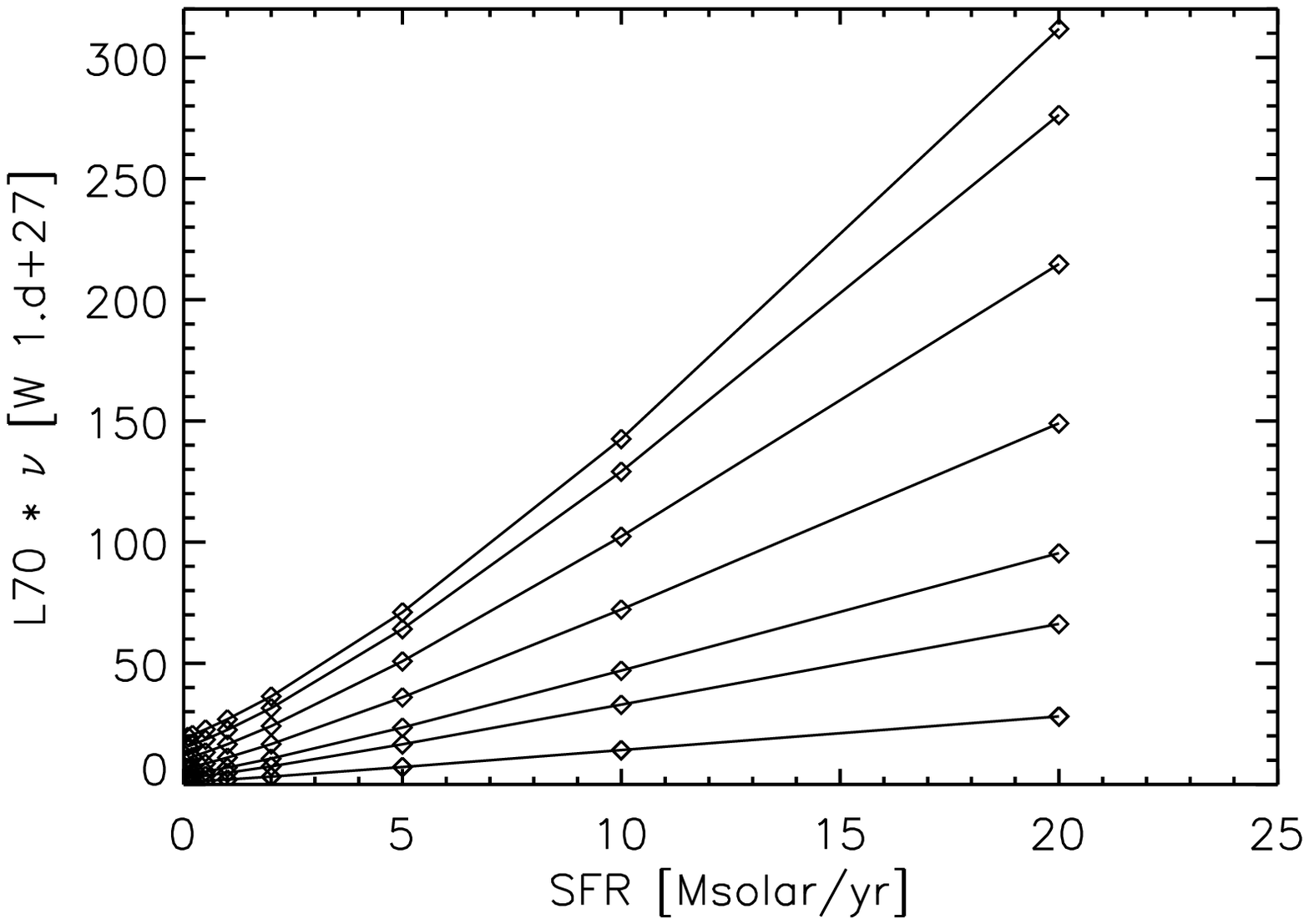}
\includegraphics[scale=0.38]{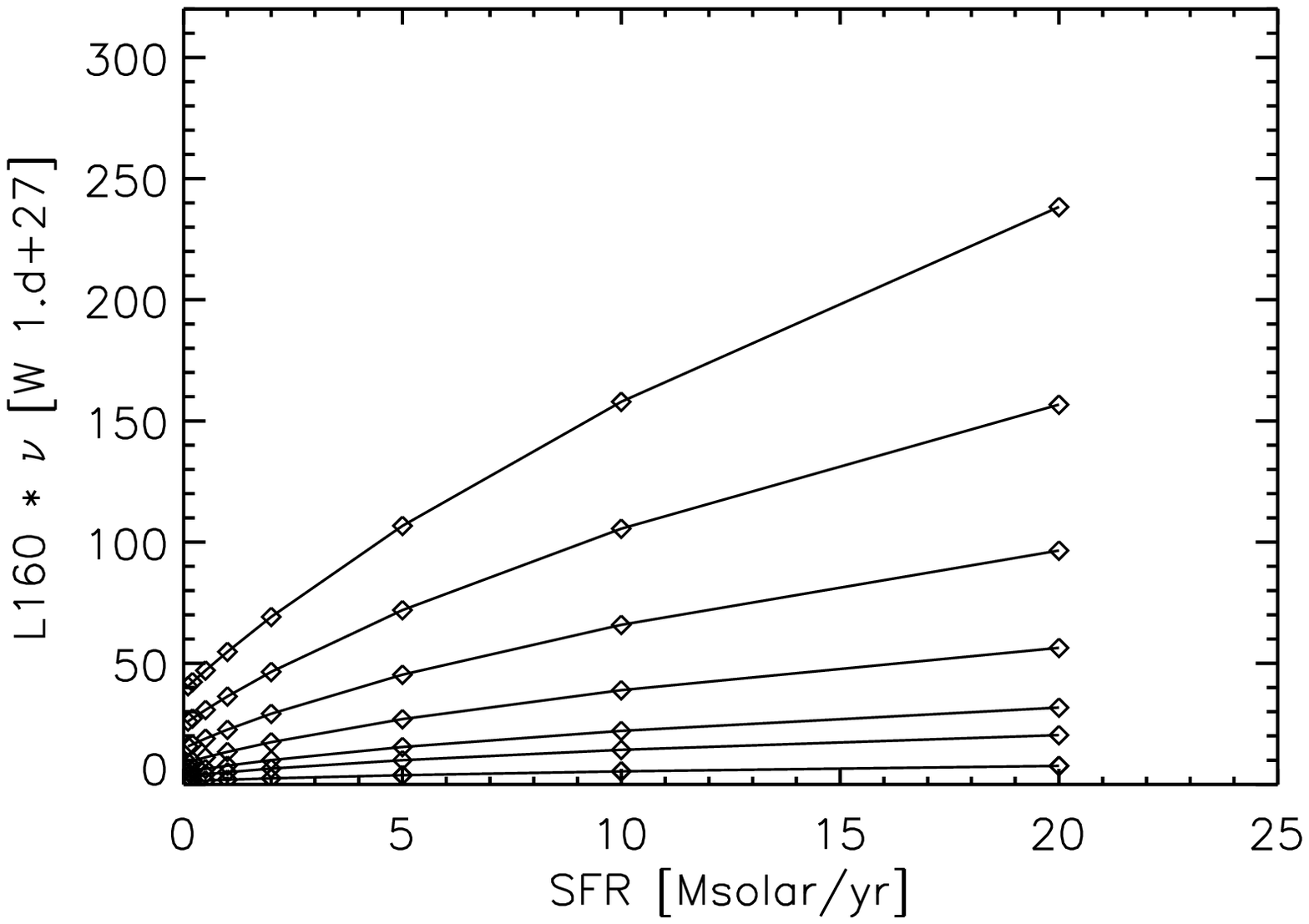}
  \caption{Predictions for the relation betwween the 70\,${\mu}$m (left) and 
160\,${\mu}$m luminosity versus SFR based on the model of Popescu et al. 
(2011). From bottom to top the curves correspond to central face-on B band
optical depth values of 0.1,0.3,0.5,1.0,2.0,4.0 and 8.0.}\label{fig:sfr_calib}
\end{figure}

SED modelling tools based on self-consistent radiative transfer calculations
can be used to predict the scatter in the SFR calibration relations, as a 
function of the main intrinsic parameters that can affect these
relations. Several of these relations have been presented. In
Fig.~\ref{fig:sfr_calib} we only show predictions for the SFR calibration based
on monochromatic FIR luminosities when the dust opacity changes. The
predictions are based on the model of Popescu et al. (2011). The figure shows a
very large scatter in the correlations. A similar large scatter is predicted 
for correlations corresponding to
various contributions coming from the old stellar populations, or for the 
clumpiness of the ISM. For the UV calibrators large scatters are also predicted
when some of the relevant parameters (viewing angle, dust opacity,
bulge-to-disk ratio and clumpiness of the ISM) vary. Overall it is apparent from
these plots that the predicted scatter in the SFR correlations due to a broad
range in parameter values is larger than observed in reality. The question
then arises of why are the SFR calibrators working, despite, for example,
the very crude dust corrections that have been so far used in the
community? A possible answer is the existence of some scaling parameters, which
do not allow a continuous variation in parameter space, in particular for dust
opacity or stellar luminosity. Recent work from Grootes
et al. (2013) proved the existence of a well-defined correlation between
dust opacity and
stellar mass density. The correlation was derived on data coming from Galaxy
and Mass Assembly (GAMA) survey (Driver et al. 2011) and the Herschel ATLAS
survey (Eales et al. 2011), in combination with the model of Popescu et
al. (2011). These finding give support to the interpretation of the existence
of fundamental physical relations that reduce the scatter in the SFR
correlations. \\

\newpage
3. {\it A word of caution}

We have identified some points where things should be treated more carefully in the 
future:
\begin{itemize}
\item The energy balance method should not be used on scales smaller than the
  scalelength of the disk, as the energy is not conserved below these
  scales. The role of long range photons in the diffuse ISM should not be
  underestimated, in particular by considering an average free path of photons
  in the disk. This is because the free path of photons firstly depends on
  radial position in the galaxy (disk), where dust opacity is known to decrease
  monotonically with radius (e.g. Boissier et al. 2004, Popescu et
  al. 2005). Secondly, there is also a vertical distribution of dust, and the
  free path of photons in vertical direction will be different from that in
  radial direction. One also needs to add the contrast between arm and
  interarm regions. Finally, the escape of photons from star-forming clouds
  is strongly anisotropic and fragmented, because of the fragmentation of the
  clouds themself. Thus, in some directions the stellar light is completely
  absorbed by dust, while there are lines of sight from which the radiation
  freely escapes in the surrounding diffuse medium. The multiple facets of the
  transfer of radiation in galaxies, including the effect of scattered light,
  means that energy balance method should not be applied on a pixel by pixel 
basis, as sometimes employed in the literature.
\item Mid-IR emission should not only be identified with ``small grain'' 
emission,  where by small grain we mean stochastically-heated grains. In fact 
in the
  range $24-60{\mu}$m most of the dust emission is powered by big grains
  heated at equilibrium temperatures by the strong radiation fields in the
  star-forming complexes.
\item The ratio between mid-IR (PAH range) emission to FIR emission cannot be 
 interpreted only in terms of relative abundances of PAH to big grains. One 
should also
  take into account the change in the colour and intensity of the radiation 
fields heating the dust, which also result in strong variations of mid-IR to
FIR emission.  
\item Do we really need to accurately know the absolute star-formation rates 
 in galaxies? Perhaps we can live with some approximations, which would be good
  enough to allow us to derive trends in galaxy populations over cosmic
  time. 
A definitive {\bf no} has been given to this suggestion. An inability to 
measure absolute SFR would severely limit our ability to constrain physical 
models of galaxies and the evolving universe. 
For example, only if we have absolute measurements of SFR will we be able to
relate measurements of SFR to measurements of gas content of galaxies in terms
of physical models predicting the amount of gas in the ISM and the efficiency
of conversion of the ISM into stars. Pavel Kroupa also gave convincing
statistical argumentation  on the need to measure accurate SFRs in the
discussion session.\\
\end{itemize}

\begin{figure}
 \includegraphics[scale=0.5]{scalelength.epsi}
\includegraphics[scale=0.5]{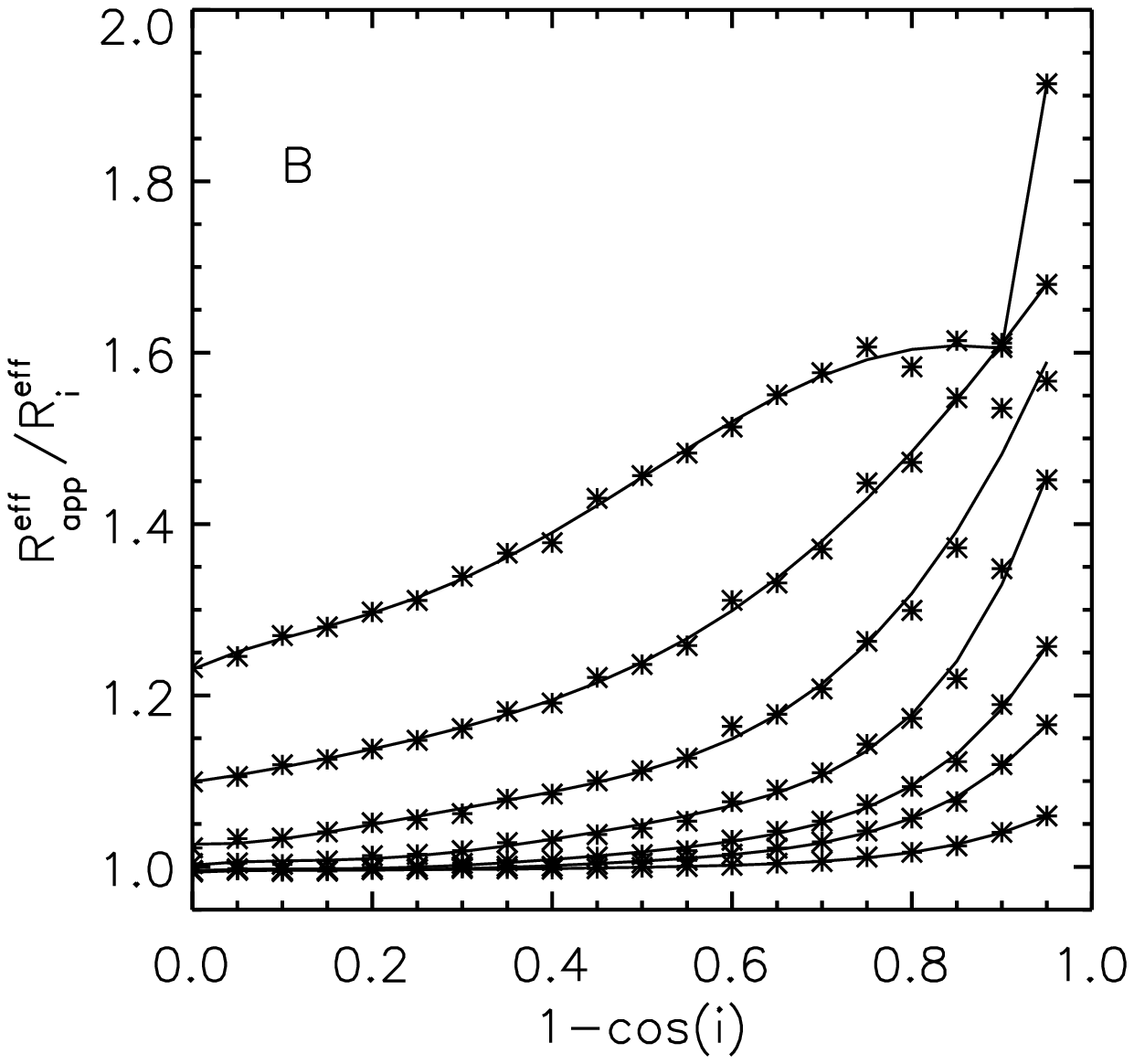}
\caption{Dust effects on the derived scalelength of disks fitted with
  exponential functions (left) and on the derived effective radius of disks
  fitted with variable index S\'{e}rsic functions, from Pastrav et
  al. (2013). Both plots are for the B band. From bottom to top curves are for
  a central face-on dust opacity in the B band of 0.1,0.3,0.5,1.0,2.0,4.0,8.0.}
\label{fig:scalelength}
\end{figure}

4. {\it What else have we learned?}

\begin{itemize}
\item Applications of self-consistently calculated model SEDs strongly rely on
scaling them according to measurements of the surface area of the
stellar disk of the  modelled galaxy, which, in turn, depends on an
accurate decomposition of the main morphological components of galaxies
as observed in the UV/optical. Bogdan Pastrav showed that the derived 
scale-sizes of stellar disks of galaxies are strongly affected by dust (see
Fig.~\ref{fig:scalelength}), and 
that a proper determination of the intrinsic distributions of stellar 
emissivity, and thus of star-formation rates, needs to self-consistently take 
into account these effects. 
\item Several panchromatic surveys, with detailed information on bulge-to-disk
  ratio, inclination and disk size are underway, making these databases ideal
  for determinations of SFR using radiative transfer models. Andreas Zezas
  presented ``The Star-Formation Reference Survey'' (Ashby et al. 2011), 
  a unique statistical sample of 369 galaxies selected to cover all
  types of star-forming galaxies in the nearby universe. The survey overlaps
  with the SDSS and NVSS areas and has GALEX, SDSS, 2MASS, Spitzer and NVSS 
  multiband photometry and planned bulge-disk decompositions of optical 
  images, which will make it ideal for self-consistent and systematic 
  determinations of SFRs. It will also asses the influence of AGN fraction and
  environment on SFR.
\item Denis Burgarella presented applications of the CIGALE SED fitting method 
on the Lyman break galaxies at $2.5<z<4$ detected in the Far-infrared with
Herschel and implication for star formation determinations at high redshift.
\item Andrew Hopkins showed results on SFRs derived from applications of the
  energy balance method on the GAMA survey. Applications of radiative transfer
  techniques to the de-reddening of GAMA galaxies by Grootes et al. (2013) were
  presented in the review of C. Popescu.\\
\end{itemize}

\end{document}